\documentclass{soups} 
\usepackage{graphicx}
\usepackage{times}

\newcommand{\specialcell}[2][c]{%
  \begin{tabular}[#1]{@{}c@{}}#2\end{tabular}}

\def\sharedaffiliation{%
\end{tabular}
\centering
\begin{tabular}{p{\linewidth}}\centering}
\begin{document}
%
\conferenceinfo{Symposium on Usable Privacy and Security
  (SOUPS)}{2017, July 12--14, 2017, Santa Clara, California.}
\CopyrightYear{2017} 

\title{Ask Me Anything: A Conversational Interface\\ to Augment Information Security Workers}
%
%
%
%
%

\numberofauthors{3} 
%
\author{
%
%
\alignauthor
Bobby Filar \\
\alignauthor
Richard J. Seymour \\
\alignauthor
Matthew Park \\
\sharedaffiliation{
\affaddr{Endgame Inc.} \\
\affaddr{3101 Wilson Blvd} \\
\affaddr{Arlington, VA 22201}
\email{\{bfilar,rseymour,mpark\}@endgame.com}
}
}
\date{26 April 2017}

\maketitle
\begin{abstract}

Security products often create more problems than they solve, drowning users in alerts without providing the context required to remediate threats. This challenge is compounded by a lack of experienced personnel and security tools with complex interfaces. These interfaces require users to become domain experts or rely on repetitive, time consuming tasks to turn this data deluge into actionable intelligence. In this paper we present Artemis, a conversational interface to endpoint detection and response (EDR) event data. Artemis leverages dialog to drive the automation of complex tasks and reduce the need to learn a structured query language. Designed to empower inexperienced and junior security workers to better understand their security environment, Artemis provides an intuitive platform to ask questions of alert data as users are guided through triage and hunt workflows. In this paper, we will discuss our user-centric design methodology, feedback from user interviews, and the design requirements generated upon completion of our study. We will also present core functionality, findings from scenario-based testing, and future research for the Artemis platform.

\end{abstract}

\section{Introduction}
Across industries organizations are faced with the growing threat of computer network attacks. Threats such as data theft, ransomware, and phishing have brought an influx of security products designed to detect and respond to intrusions and anomalies on a network. However, for these platforms to be employed successfully, organizations require a workforce capable of not only having a subject matter expertise in information security, but also a deep understanding of the very platforms designed to augment their day-to-day operations. 

The construction of a viable security operations center (SOC) has proven to be the one of the biggest issues in information security.\cite{zadelhoff_2017}  The primary mission of a SOC is network defense. It is a difficult and demanding task requiring analysts to manually search through a deluge of alert and log data to identify anomalies. Experienced analysts rely on both personal and organizational methodologies to: 1) formulate a hypothesis; 2) gather supporting evidence and 3) remediate a threat. 

One of the key challenges in creating a SOC is the lack of skilled professionals. Currently, the security industry faces a workforce shortage with the demand of skilled professionals far outpacing the pool of available applicants.  Estimates of this deficit range from 1-2 million workers by 2019.\cite{csx2015} Multiple factors contribute to the shortage, but two that stand out are a lack of subject matter experience and a lack of educational tools for potential workers. In SOC environments, this is compounded by non-intuitive interfaces and complex query languages found within products that force users to be experts in both security and the platform itself. Current approaches to tackle this problem focus on augmenting analysts through standardized analytic processes, such as collaboration and information sharing, and less on training and educating inexperienced analysts. Assistive technologies, such as conversational interfaces, are rarely included in security platforms and represent a missed opportunity to develop expertise holistically.

In this paper, we present preliminary research from studying diverse user groups within security organizations. We detail the roles, behaviors, and workflows employed during day-to-day operations we uncovered throughout the research process. We highlight common challenges shared across security teams such as alert fatigue, data deluge, and complex user interfaces. From this study we present Artemis, a conversational interface to the Endgame Endpoint Detection and Response platform. Artemis was developed in response to the data collected during our study by employing a user-centered design approach. The results of this research yielded invaluable insights we leveraged into development requirements for our conversational interface that will be discussed in-depth in the following sections. Finally, we share our findings from testing the initial implementation of Artemis and provide recommendations for future iterations.

\subsection{Related Work}

Previous research in augmenting the security workforce has centered on collaboration through co-locating analyst environments with data sources\cite{staheli2016collaborative}, reducing alert fatigue\cite{assante2011enhancing}, and using simulations to improve training \cite{herr2015video,baker2016striving}. Most of this research focuses on process, rather than tools designed to empower analysts. 

Conversational interfaces or dialog systems have been successfully employed in a variety of domains. These intelligent assistants can provide ``best practice" guidance and recommended paths to desired actions within an intuitive, natural language interface. In \cite{holzinger2012knowledge} researchers applied NLP to provide an interface enabling users to find and recognize previously unknown, but useful information. This required domain expertise, thus making the identification of important, but previously unknown, information significantly easier.  Researchers in \cite{celikyilmaz2014resolving,crook2016task} discuss the benefits of personal digital assistants (e.g. Cortana) to carry out Question-Answer tasks via natural language queries against the web or knowledge bases. In information security, domain researchers in \cite{palmer2016cognitive} introduce an intelligent assistant that leverages IBM Watson natural language understanding software for analyzing unstructured security data.

\section{User-Centered Design}
The User-Centered Design approach\cite{usability.gov_2017} focuses on three primary phases: 1) Discovery; 2) Concepting and 3) Prototyping and User Testing. We designed our Discovery phase to better understand our target user base. By capturing team dynamics and organizational workflows we came away better equipped to generate product requirements. We began this work early in the research and development process starting with user interviews across a variety of use cases, including red vs. blue teams, where attackers (red team) attempt to infiltrate a computer network defended by a blue team in a mock scenario.

\begin{table*}
\centering
\caption{Breakdown of Security Team Data Collection.} 
\label{table:secteam}
\begin{tabular}{|c|c|c|c|} \hline
Group&Team Type&Discovery Environment&Collection Method\\ \hline
A&Traditional SOC&Day-to-Day Use&User Interviews\\ \hline
B&Novice Training Team&Mock Scenario&\specialcell{Side-by-side monitoring,\\ Retrospective \&\\User Interviews\\w/ guerrilla methods}\\ \hline
C&Red v. Blue (internal)&Mock Scenario&\specialcell{Side-by-side monitoring,\\Retrospective \&\\User Interviews\\w/ guerrilla methods}\\ \hline
D&Traditional SOC \& Security Consultants&Day-to-Day Use&User Testing w/ guerrilla methods\\
\hline\end{tabular}%
\end{table*}

These efforts were overseen by personnel from Endgame Product, UX, and R\&D groups and carried out internally and externally. A combination of mock scenarios and user interviews were conducted with users at every level of a security team. The teams selected for observation were carefully chosen to represent the wide array of blue teams found in a typical SOC environment. Questions and scenarios were modified to reflect the diversity of the teams, verticals, and sectors they represented. From data collection methods used, a set of four users archetypes were identified and their roles and responsibilities helped shape the design of Artemis. The teams selected were as followed:
 
\subsubsection{Traditional SOC}
The Traditional SOC was a veteran security team that is still currently employed in the commercial space. The participants from this team included four highly trained SOC professionals, two moderately trained SOC analysts, and one manager overseeing the team's operation.  They were familiar with a multitude of security analysis tools, including Endgame's platform.  
 
\subsubsection{Novice Training Team}
The Novice Training Team was a hand-picked, relatively inexperienced security team, which currently operates in the federal space. The participants from this team included ten new SOC professionals, and two highly trained SOC leaders.  While the two experienced analysts were familiar with many security analysis tools, the other ten participants had a base familiarity with two tools, one of which was the Endgame Platform.
 
\subsubsection{Red v. Blue (internal)}
The Red v. Blue (internal) was a mixed group of security professionals, with backgrounds mostly from the federal space. The participants from this group were split into two teams. The blue team consisted of four highly trained security professionals, two moderately trained analysts, and two inexperienced analysts. The red team consisted of three extremely proficient security professionals. Participants had a mixed range of knowledge on the Endgame platform, as well as other security analysis tools.
 
\subsubsection{Traditional SOC \& Security Consultants}
The Traditional SOC \& Security Consultants was an ever-rotating cast of different security professional types seen from both the federal and commercial space. The participants included a wide array of security analysts and were typically interviewed in teams of one or two on one-week intervals. There was usually little knowledge of the Endgame Platform, but participants were well versed in other security analysis tools.

\subsection{Data Collection}
We sought to employ a variety of collection methods on the diverse set of user groups during the discovery process. (Table \ref{table:secteam}) Most participating individuals were interviewed in one-on-one settings against a set similar questions. Interviews consisted of 12 predefined questions based upon team environment and 4 questions dependent on the user's role within a security organization. These questions sought to elicit user views on the current state of the UI, viability of the contextual information provided on alert views, and recommended capabilities, functionality, and data sources. Additional insight was gained during operational multi-day exercises where UX, product, and research personnel were embedded as monitors and captured workflows and user behavior in a simulated environment.
 
Other collection methods involved multi-hour retrospectives where teams, consisting of managers and analysts (experienced and junior), were given the opportunity to respond to questions about the platform, their workflows, and the opportunity to provide user feedback. Finally, user testing with guerrilla methods was situationally employed during the multi-day operational exercises and a long-term testing phase with a mid-sized security team. Guerrilla usability testing allows researchers to rapidly assess user opinion on a feature, but having the user provide thoughts while actively using the product ``in the wild."\cite{unger_warfel_2011} The results are not quantitative, but rather qualitative, uncovered through conversation. Guerrilla methods allowed us to gain quick feedback on incremental changes about the product by approach various members of security teams. Guerrilla methods require a limited time commitment and does not impede with security workers day-to-day responsibilities. Moreover, this methodology adequately captured changes in opinion on core features of our platform over an extended testing period. 
 
\textbf{The Traditional SOC Team} The Traditional SOC Team was interviewed in two distinct sessions. First, the group participated in a four hour informal retrospective where members first identified their roles and were given the opportunity to ask questions and provide feedback with a loosely set agenda. The Team was then split into individual formal sessions where they were presented with the 16 predefined questions. Finding were discussed internally and broken out into a mental model. All interviews were recorded, and transcribed. 
 
\textbf{Novice Training Team and Red vs. Blue} Since both the Novice Training Team and the Red vs. Blue (internal) Team matched in general size, background space, and field experience, they each participated in the same multi-day operational exercise. Participants formed a 10-12 person blue team, where they were given the operational task to search out red team attackers and remediate their mock environment. Our red team consisted of highly trained security professionals who were tasked to invade the mock environment. Both exercises were intensely monitored and throughout the mock scenario UX personnel engaged all members of the blue team in quick 15 minute one-on-one guerilla testing. At the end of the exercise both groups engaged in a 1 to 2 hour retrospective.  Field notes were taken throughout the entire exercise.
\textbf{Traditional SOC \& Security Consultants} The Traditional SOC \& Security Consultants Team was engaged over the course of several months in quick guerilla testing sessions. UX and Product personnel approached 1 to 2 members of the team to ask questions around the product, observe them perform specific tasks while using the product, and asked about their experience. These sessions lasted 30 minutes, were weekly to biweekly and were recorded using field notes.
 
Overall, our collection methodology provided a rich data source to generate standard security team workforce roles and set requirements to begin the design of our conversational agent.

\subsection{Workforce Roles}
Upon completion of the of the discovery phase, we were able to aggregate collected data to clearly define four prominent roles in most security organizations: SOC Manager, Tier 1, Tier 3, and Incident Responder.
\subsubsection{SOC Manager}
SOC Managers are often skilled security practitioners, but are not necessarily subject matter experts. They have extensive management experience and oversee day-to-day team operations. This role leverages ticketing and workflow tools, such as SIEM dashboards, to generate automated reports.  They are responsible for assigning and prioritizing investigation and triage efforts. They manage and author security playbooks, create analyst development plans, and brief C-level executive teams on the current state of the network. 
\subsubsection{Incident Responder}
An incident responder or forensic investigator is a security expert specializing in endpoint discovery platforms and sophisticated investigation tools. Most comfortable with command lines and scripting languages, an incident responder often bypasses the Graphical User Interface (GUI) altogether to use third-party APIs which allows for rapid data collection to determine the origin and extent of a possible breach. Although tasked with preparing response plans for incidents, incident responders typically do not have the authority to respond with a decision from the network owner.
\subsubsection{Tier 3 Analyst}
Tier 3 analysts are expected to intimately understand network and platform architecture, investigate escalated alerts, determine root causes, and remediate problems.  They are depended on as domain experts within the security team and are more comfortable working through the command line. They prefer a wealth of data being pulled back through an endpoint platform as they know exactly what they should be looking for and what steps are needed to prepare a plan of action. They are often critical of endpoint platforms' GUI, which they view as restricting them from quickly carrying out tasks and workflows during an investigation.
\subsubsection{Tier 1 Analyst}
Tier 1 analysts have limited prior experience (approx. 1 year) in the information security space. Professional experience is gained through professional training, often a 1-2 week course, and through on-the-job training. A typical Tier 1 will have a basic understanding of their organization's network and the security platform they employ.

Their primary role is the \textit{first line of defense} on a security team, responsible for triaging alerts and determining if escalation (passing to higher tiered analyst) is required.  An exorbitant amount of data, which is often not relevant to determine severity of an alert, frustrates these users as they look for a select amount of data to determine escalation.  These users primarily rely on a platform's GUI. Adding new syntax from the information security space or the organization's platform further complicates this user's understanding of what warrants escalation. Additional responsibilities for Tier 1 include monitoring, review, and escalation of new alerts on the platform.\\

\subsection{Findings}
Throughout the data collection process we discovered commonalities across the teams interviewed. While security workforce roles provided valuable insight, it was the descriptions of analyst day-to-day workflow that proved most useful.

\textbf{Contextual Information} (Tier 1): Many analysts described a lack of understanding of how best to interact with various features within the security platform. Sample feedback included:
\begin{itemize}
\setlength{\itemsep}{1pt}
\setlength{\parskip}{1pt}
\setlength{\parsep}{1pt}  
    \item \textit{What data returns from a persistence hunt?}
    \item \textit{What does this alert type mean?}
    \item \textit{How do I whitelist an alert?}
    \item \textit{Now that I created an investigation, where am I going?}
\end{itemize}

Generally speaking, this takeaway leads to unnatural or bad habits.  Because users don't know what features do, or why they can't easily perform some task - they go through a grueling process of repeatedly accomplishing the same task in many, many steps.

\textbf{Pivoting} (Tier 1/Tier 3): Another common theme among analysts was pivoting throughout the platform.  This observation was closely tied to context,. Most users wanted an easier way to go from an alert or data point back into a data store to extract supplemental or complementary information. This step is critical for analysts to develop a fully formed hypothesis about an event.  A natural, intuitive workflow for pivoting within the platform simply did not exist. Because of this, the user often opened multiple tabs to access core functionality and data, and yield supporting facts necessary to remediate an alert. 
\subsubsection{Day in Life}
When a SOC analyst starts their shift, they first participate in the shift handover from the analysts on the previous shift. Here they get a briefing on current ongoing investigations or open alerts, ticket numbers associated with those alerts, who is assigned to the investigations, and anything that needs attention.

They then monitor a SIEM, an assigned endpoint UI dashboard, or email and wait for a security event to occur. Typically that doesn't take very long - with the amount of tools generating alerts in a typical SOC environment (especially those monitoring large networks) - getting alerts is not the problem. \textit{Determining which alerts to focus on is the problem.} The analyst is typically in reactive mode - where they respond to alerts as they come in, quickly identifying the high priority alerts. 
 
Typically, Tier 1 Analysts will have little to no authority to take immediate action on suspected malicious security events and will instead move the alert up the SOC chain. When escalating they will create a case/investigation/incident and assign that case/investigation/incident to the SOC investigator (or Tier 3 analysis).  Both the SOC Investigator and the Tier 3 Analyst will take further steps in verifying the anomalous event, and will take the proper response in remediating the alert. While the Tier 3 Analyst will also sift through a SIEM alongside the Tier 1 Analyst, SOC Investigators will often times only work on escalated alerts. 
 
At the end of the shift, all levels of analysts needs to prepare a report of the alerts triaged, what was resolved, and what is still open in order to handover the activities to the oncoming shift.  Larger reports depicting alert and investigation trends are generated for a SOC Manager on a daily or weekly basis. The SOC Manager will use these reports to focus in on key metadata in the coming weeks, determine the SOC shift schedule and build a custom summary report for the organization's executive level. (Table \ref{table:tasktime})

\begin{table}
\centering
\caption{The Breakdown of Time Spent on a Specific Task.}
\label{table:tasktime}
\begin{tabular}{|l|l|} \hline
    Task& \% Time Spent\\ \hline
Shift Handover& 12.5\\ \hline
Report Writing& 12.5\\ \hline
Incident Creation& 25\\ \hline
Alert Triage& 50\\
\hline\end{tabular}
\end{table}

\subsubsection{Design Requirements}
On top of defining security team work roles, we leveraged collected data to develop a set of design requirements that sought to meet the demands of and eliminate common frustrations across security workers. At its core, we needed a solution to become the interface to network endpoint data, providing an ability to generate natural language queries, and perform complex workflows. At a minimum, the solution must:

\begin{enumerate}
\setlength{\itemsep}{1pt}
\setlength{\parskip}{1pt}
\setlength{\parsep}{1pt}  
\item Eliminate query syntax via natural language
\item Educate users on platform features
\item Provide context-driven alert triage
\item Recommend next steps
\item Expedite focused collection
\end{enumerate}

These requirements and additional functionality resulted in Artemis, a conversational interface,  described in the following sections.

\section{Artemis}

\begin{figure}
\centering
\includegraphics[width=3.0in]{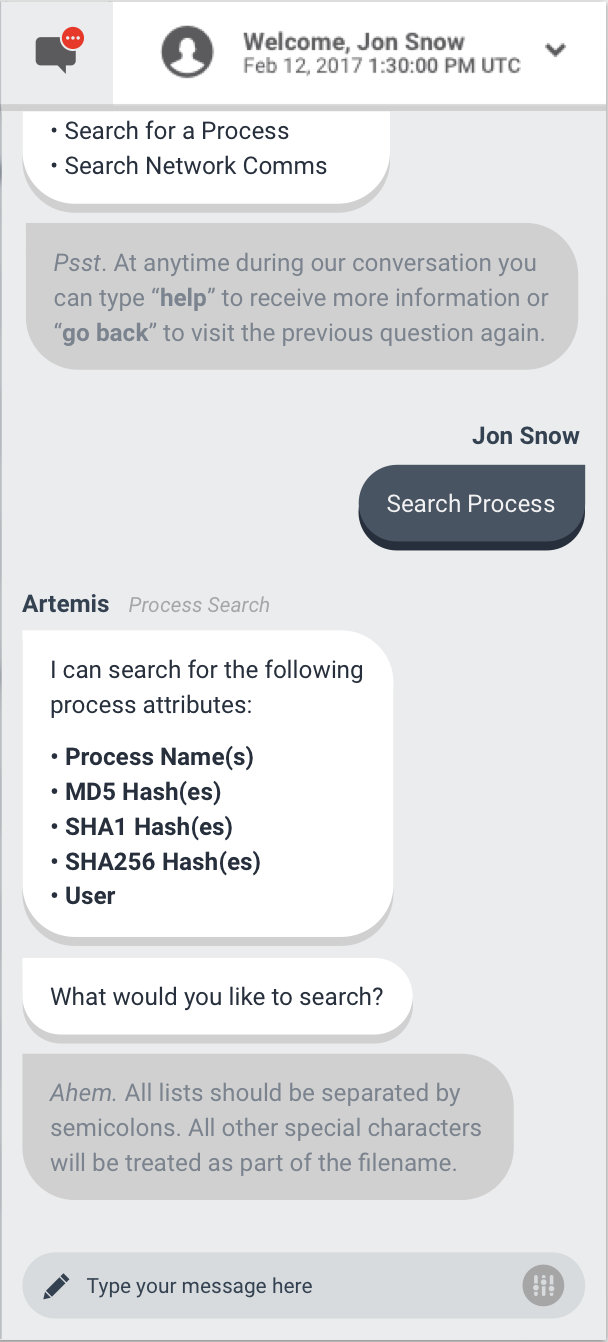}
\caption{Primary pane of Artemis Chat Interface}
\end{figure}

A conversational assistant met the demands of multiple users and use cases. Artemis was our name for the suite of technologies that presents a chat interface to the user. (figure 1) It is more than just a bot, but rather enabled multiple advances in user experience and functionality. We combined UX research and data science to create a user-centric design to ease users' pain points as described below.\\

\subsection{Conversation-driven Queries}
To oversimplify  Artemis, a user asks a question and Artemis returns the answer. The more detailed view is that the user crafts a natural language query across endpoints and Artemis sends that query to those endpoints and then formats the data returned. The goal was to allow users to move away from syntax driven queries (such as SQL or Lucene query syntax) and more towards semantically driven queries. In other words, the ideal interaction would never require the user to translate what they mean from English into the syntax of the computer. Understanding the entirety of the written word requires contextual information, but limiting the understood vocabulary to the information security world allows for quicker understanding. The bot doesn't need to disambiguate between process as an abstract concept and process as a running program on a computer, since it is always the latter.

Parsing natural language sentences to guide less experienced tier 1 users through their initial interactions can speed up their workflow. For Tier 3, though, we can't let the ease of use become a hindrance to their speed of interaction. User feedback showed that typing full sentences could get tiresome search after search. The bot luckily has no programmed requirement that the users communicate in full sentences. Once the intent is given, the rest of the parameters can be added in with very little extraneous verbiage.\\ 
\subsubsection{Language Understanding}
Supporting communication with the end user may seem magical at first, but the code behind it performs two basic functions which are enabled by recent advancements in open natural language understanding (NLU) research. The user's utterance, or text input, is passed into an entity extractor, which categorizes the words and phrases into entities. In NLU this tagging process is known as Named Entity Recognition (NER). We then replace the entities with a generic name through a process we call redaction. This allows the next step of intent classification to model a reduced vocabulary.

\begin{figure}
\centering
\includegraphics[width=3.3in]{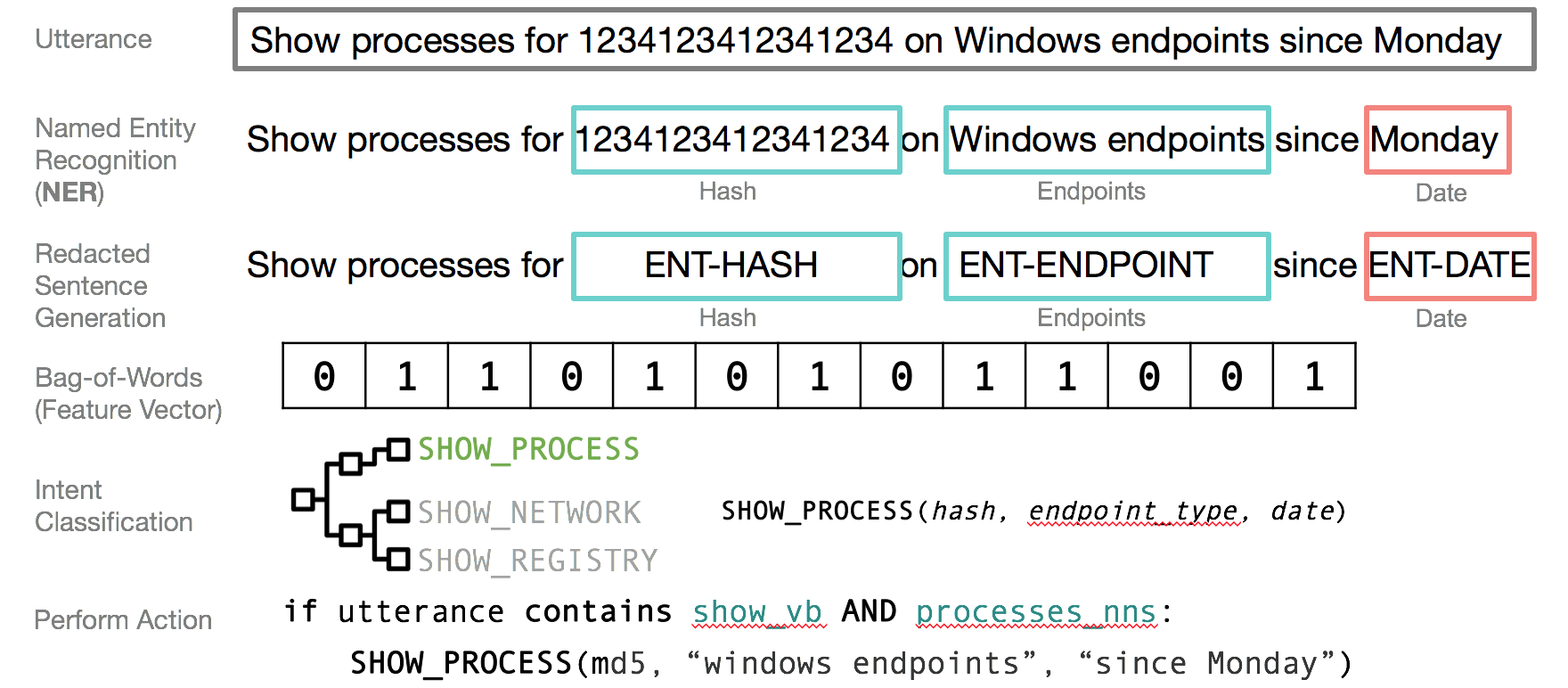}
\caption{NLU Pipeline on a sample Artemis query}
\end{figure}

The second and final step in the text processing pipeline is classifying the intent of the user's utterance. (figure 2)  The intent corresponds roughly to what the user wants to do and the entities correspond to what the user wants to do it with.  The intent classifier works by analyzing the entire redacted sentence and comparing the features to models trained on a set of manually classified sentences.
\subsubsection{Fill in Missing Information}
The two stage language understanding pipeline described works for most cases, but the models can fail to classify certain utterances or entities. We try to avoid this type of classification error through robust training sets and rigorous testing. Admittedly the engine can do very little if it fails to recognize any intent or entity whatsoever. In our testing, if the user doesn't enter a recognizable intent we can learn from this data to improve our models. In deployed usage, though, the user still wants to achieve their goal regardless of Artemis' understanding.  If the user phrases something such that our intent classification fails, we offer helpful dialog to guide the user and elicit intent.

A more common missing information situation is when an intent has been classified but the user hasn't filled in all of the entities needed to run the task.  The tasks available to the user have schemas which dictate the required and optional parameters, and correspond to different entity types.  Selecting and asking for the remaining entities for a given intent is as simple as comparing the set of given entities to the set of needed/optional entities and asking for the difference in a generated sentence.

\subsubsection{Misunderstanding and Disambiguation}
Other than intent classification failure and entity extraction failure, the most pernicious problem is misunderstanding. Classifying the intent as the wrong intent is arguably worse than classifying it as no intent. For example if a user wrote \textit{``search for a process"} and got into the \textit{``kill\_process"} intent, the results could be very damaging. We avoid this by limiting the words that we train with for more dangerous tasks, lowering the likelihood of misclassification. This additional safeguard against dangerous tasks, unit testing, and quality assurance, plus our confirmation step, protect users from this possibility.

The more likely situation is that the entities are extracted improperly.  For example, filenames are limited in characters, but still have so many possible combinations that many domain names, IP addresses, hashes, and usernames are actually valid filenames as well.  How can a user or a computer know if \textit{``command.com"} refers to a filename or a website? Without context that differentiation is impossible and it wouldn't help the user to guess. In this case of multiple valid slots open in the intent for multiple possible entity extractions, we must disambiguate with the user. This slows workflow, so we support hinting or tagging the words such that \textit{``file command.com"} would definitively be extracted as a file name.
\subsection{Precision Search}
In 2008 the MapReduce paper \cite{dean2008mapreduce} showed that processing big data could be done efficiently by distributing the computation to the computers that held the right shards of data. In information security, data from endpoints is often retrieved to a central location and processed.  In fact, the data is sometimes taken from a set of endpoints, then stored in a cluster for centralized processing.  In our work with Artemis, we wanted to leverage the computing power of the endpoints in a similar way to MapReduce.  We treat the endpoints as the worker nodes, responsible for their own data, then map a function across the endpoints to collect the relevant data for the user.  That way only the relevant data goes across the wire, limiting after the fact data sifting.
\subsection{Context-Aware Dialog}
Artemis' ability to infer meaning is imperative to the workflow. 

\textit{\textbf{Conversation with a Participant during training:}\\
``It's great you have all this information here. But here, ok. Look at this PID (they are pointing to a random PID in a process hunt on the endpoint page). What can i exactly do with this? Can I run a process survey on this? Can I search on this to get relevant information to this survey? What are the hunts I can do with this? I want to have a way I can do this from here, without spending the time searching around.. building information around this PID."}

To reduce the amount of clicks and eliminate the need to perform copy/paste operations on long, complex search strings (i.e. file hashes or registry keys), Artemis is able to leverage the data currently in view to allow the user to express query terms generically. For example, if a user is currently viewing a malware alert, which provides information such as a file hash and name, they do not need to copy or type that metadata into Artemis. Instead, the user may express their search more generally using an utterance like \textit{``does this hash show up anywhere else in my network?"} or \textit{``search process data for this hash"}. This functionality should aid in giving experienced and inexperienced users more time in the chat window and executing workflows, and less time clicking to perform copy/paste operations.
\subsection{Card-based Results View}
The successful completion of a natural language query launches an investigation to endpoint machines. The results of that action are presented in a card-based view for quick digestion. Cards are ideal for communicating quick stories about each returned results. Each card summarizes essential data, including information about the query, as well as supplemental metadata that can be used to pivot to the next operation. The cards employ color-coding and iconography to highlight cards that may require immediate attention. (figure 3) These cards have been tagged by an enrichment pipeline all returned data passes through to determine if query responses contain any anomalous information (e.g. high likelihood of malware). 

\begin{figure}
    \centering
    \includegraphics[width=3.3in]{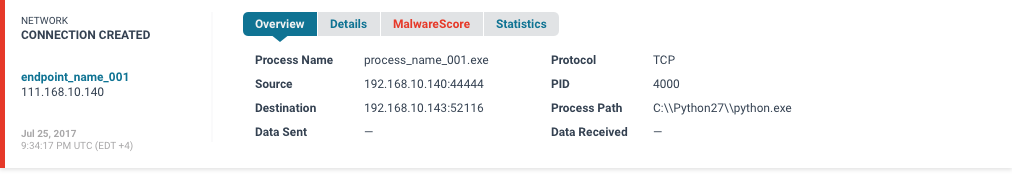}
    \includegraphics[width=3.3in]{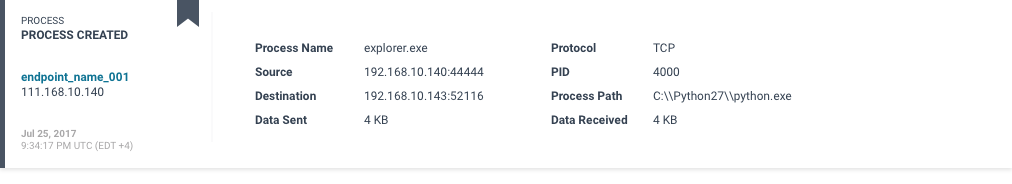}
    \caption{(top) Red coloring highlights enriched record; (bottom) visual indicator shows user the starting point of a process lineage task}
\end{figure}

\section{Results}
Our goal was to capture the challenges facing security teams and understand ideal day-to-day workflows for experienced and inexperienced security workers. The initial implementation of Artemis is a byproduct of those findings.

After integrating Artemis into the Endgame platform we conducted follow up interviews and user testing. User stated they spent a minimal amount of time understanding how best to interact with the tool. Tier 1 analysts highlighted the utility in turn-based conversations to construct complex queries and build investigations. Several Tier 3 analysts stated they were skeptical of a ``chat" interface, but after extended use began to treat it as a command line to launch investigations and perform queries across network. Tier 3 and incident responders maintained that for mass ingest of data (e.g. searching 100s of file hashes from threat intel feeds) they would prefer to script against our API. Both sets of users stated they would continue to use the tool and recommended additional capabilities to improve day-to-day workflows.

We monitored security worker teams in small scenario-based simulations that required use of Artemis to better capture how workers interacted with the tool. We present the following categories that capture the introduction of Artemis into their workflow.

\subsection{Create Efficient Alternative Interactions}
Offering alternative interactions with data tailored to user needs can improve interaction efficiency for a variety of users. Our platform is built around a GUI for displaying and interacting with security information provided by sensors deployed on endpoints.  While this is immediately usable by most information security professionals, those used to the command line may find it too point and click intensive.  Keyboard shortcuts may sometimes make users more efficient, they can be too hard to recall \cite{Miller:2011:CCH:1978942.1979351,tak2013satisficing}. With a conversational interface we are able to reduce the memory requirements of keyboard shortcuts by allowing multiple paths to success, while ideally reducing the search and click latency of mouse driven interfaces. The ability of Artemis to chat with the user removes some of the pain of a command line interface, where an incomplete command fails silently, our bot will ask for the information necessary to complete a task. 
\subsection{Reduce Focus Shifts}
Flipping back and forth between web pages, tabs, modals, can be tiresome for a user, especially as the data changes within the components.  Some observations from our user interviews explain the focus difficulties faced.  

An analyst would have 3 tabs open at all times. When noticed, the room was quickly surveyed to see who was working in a similar way. On average, 2-3 tabs were opened on everyone's computer to manage the data and ease workflow.

\textbf{Witnessed Ex 1:} If a user was interested in a subset of returns of an investigation and wanted to task those endpoints:
\begin{enumerate}
\setlength{\itemsep}{1pt}
\setlength{\parskip}{1pt}
\setlength{\parsep}{1pt}  
    \item The Investigation Detail page was opened to view the endpoints in question.
    \item The Endpoint List page was opened to filter through the data to select those particular endpoints.
\end{enumerate}
\textbf{Witnessed Ex 2:} When a user wanted to compile similar alerts (within different endpoints) and link them to an investigation, they had to:
\begin{enumerate}
\setlength{\itemsep}{1pt}
\setlength{\parskip}{1pt}
\setlength{\parsep}{1pt}  
    \item Keep one tab open to cycle through alerts to copy/paste endpoint names in a table (MS Excel).
    \item Have the endpoint table open to then filter those results to find the endpoints, to task the alerts in question.
\end{enumerate}

With a chat pinned to the window the user can see the data retrieved from their chat and ask further relevant questions. This reduction in focus shift allows users to avoid context switching penalty while maintaining their analytic flow. 

\subsection{Personal Expression to Canonical Search}
Each user or team may have a different vocabulary describing the world of infosec. A conversational interface should allow for synonymous language to be used without concern for proper keywords and syntax.  We call the words and phrases in a user's vocabulary, mainly nouns, \textit{entities}, a term taken from NLP research. The conversational interface classifies such entities in an utterance into a known canonical set. For example, users are free to express themselves in such a way that if they like to call machines ``boxes" or boxes ``machines" or ``computers" the bot can classify them all as ``endpoints."  This classification allows the bot to enact a version of the Robustness Principle (Postel's Law),  ``Be conservative in what you send, be liberal in what you accept".  This allows for a shared canonical language to be used for saved searches and within the product, without limiting users to our chosen terminology or forcing them to recall the exact phrasing they need to use.
\section{Future Work}
The current implementation addresses major challenges analysts face in search and discovery tasks, provides a platform to implement workflows, and presents an interface that can educate inexperienced users. Remaining challenges lie in the platform's ability to provide simple, intuitive collaboration to users and better natural language understanding. 
\subsection{Collaboration}
The current implementation addresses major challenges to analysts face in search and discovery tasks, provides a platform to implement workflows, and presents an interface that can educate inexperienced users. Remaining challenges lie in the platform's ability to provide simple, intuitive collaboration to users. Users have expressed the desire to share investigations with other members of the team. 

\textit{\textbf{Conversation with a Participant during training:} \\
``Ok, I've been assigned this set of alerts.. but why? What should I be looking for? Does the person assigning me this have insight to where I should start my hunting? How do I even notify that user that this task has been completed once I have resolved the issue?"}

Our proposed approach focuses on allowing the users to ``at" (e.g. like @ on Twitter) each other or all members of team with a completed investigation. Recipients of these ``@s" will see a broadcast message appear in their chat window alerting them to check the investigation page for a newly shared event. The first tier of collaboration should satisfy most requirements in the near-term and provide a foundation for actual chat integration.
\subsection{Chat Integration}
While our platform is usable through a point and click interface, our power users can already integrate their own tooling with our exposed APIs.  The API of Artemis is also exposed to our users under authentication.  This allows for experimental integration with other chat platforms and modalities, such as Slack, HipChat and Jabber to Alexa and Cortana.  We have yet to create these integrations ourselves, although we are happy to open it up to our users.
\subsection{Improve via Active Learning}
Active learning presents the end user an opportunity to evaluate how well the natural language pipeline performed in understanding an input. UX considerations will be necessary to determine the most advantageous approach to collect this data with possible options including: 1) Artemis asking the user how well it did; 2) Providing a clickable ``thumbs up/down" indicator; or 3) Upon completion of a task asking the user to grade performance on a 1-10 scale. Collected feedback will be applied in a semi-supervised learning algorithm to augment future classification models.
\section{Conclusion}
As the need for highly skilled security workers continues to rise, it is imperative to provide capabilities designed to augment, not hinder, the development of current and future analysts. Through the collaborative effort of UX, data science, and domain experts, security tools can be developed to support the diverse user base they serve. Artemis and conversational interfaces in general offer one possible solution to creating an accessible interface for new users while simultaneously offering faster, more complex interactions for experienced users.

Technology built upon user-centered design can augment the current workforce and reduce the barrier of entry for future security workers. Similar to our user-focused research in developing Artemis, the security industry needs to integrate data automation within user-friendly interfaces that meet the workflows of analysts tasked with defending against the growing range of attacks. Not only does this augment the current workforce, but it also will make security more accessible to a broader workforce in the coming years.

\bibliographystyle{unsrt}
\bibliography{sigproc}  

\begin{thebibliography}{10}

\bibitem{zadelhoff_2017}
Marc~van Zadelhoff.
\newblock Cybersecurity has a serious talent shortage. here's how to fix it,
  May 2017.

\bibitem{csx2015}
ISACA CSX.
\newblock Global cybersecurity status report, Jan 2015.

\bibitem{staheli2016collaborative}
Diane Staheli, Vincent Mancuso, Raul Harnasch, Cody Fulcher, Madeline
  Chmielinski, Adam Kearns, Stephen Kelly, and Era Vuksani.
\newblock Collaborative data analysis and discovery for cyber security.
\newblock In {\em Twelfth Symposium on Usable Privacy and Security (SOUPS
  2016)}. USENIX Association, 2016.

\bibitem{assante2011enhancing}
Michael~J Assante and David~H Tobey.
\newblock Enhancing the cybersecurity workforce.
\newblock {\em IT professional}, 13(1):12--15, 2011.

\bibitem{herr2015video}
Christopher Herr and Dennis Allen.
\newblock Video games as a training tool to prepare the next generation of
  cyber warriors.
\newblock In {\em Proceedings of the 2015 ACM SIGMIS Conference on Computers
  and People Research}, pages 23--29. ACM, 2015.

\bibitem{baker2016striving}
Marie Baker.
\newblock Striving for effective cyber workforce development, 2016.

\bibitem{holzinger2012knowledge}
Andreas Holzinger.
\newblock On knowledge discovery and interactive intelligent visualization of
  biomedical data-challenges in human-computer interaction \& biomedical
  informatics.
\newblock {\em DATA}, 2012:9--20, 2012.

\bibitem{celikyilmaz2014resolving}
Asli Celikyilmaz, Zhaleh Feizollahi, Dilek~Z Hakkani-T{\"u}r, and Ruhi
  Sarikaya.
\newblock Resolving referring expressions in conversational dialogs for natural
  user interfaces.
\newblock In {\em EMNLP}, pages 2094--2104, 2014.

\bibitem{crook2016task}
PA~Crook, A~Marin, V~Agarwal, K~Aggarwal, T~Anastasakos, R~Bikkula, D~Boies,
  A~Celikyilmaz, S~Chandramohan, Z~Feizollahi, et~al.
\newblock Task completion platform: A self-serve multi-domain goal oriented
  dialogue platform.
\newblock {\em NAACL HLT 2016}, page~47, 2016.

\bibitem{palmer2016cognitive}
Charles Palmer, Lee~A Angelelli, Jeb Linton, Harmeet Singh, and Michael
  Muresan.
\newblock Cognitive cyber security assistants-computationally deriving cyber
  intelligence and course of actions.
\newblock In {\em 2016 AAAI Fall Symposium Series}, 2016.

\bibitem{usability.gov_2017}
Department of~Health and Human Services.
\newblock User-centered design basics, Apr 2017.

\bibitem{unger_warfel_2011}
Russ Unger and Todd Warfel.
\newblock Getting guerrilla with it, Feb 2011.

\bibitem{dean2008mapreduce}
Jeffrey Dean and Sanjay Ghemawat.
\newblock Mapreduce: simplified data processing on large clusters.
\newblock {\em Communications of the ACM}, 51(1):107--113, 2008.

\bibitem{Miller:2011:CCH:1978942.1979351}
Craig~S. Miller, Svetlin Denkov, and Richard~C. Omanson.
\newblock Categorization costs for hierarchical keyboard commands.
\newblock In {\em Proceedings of the SIGCHI Conference on Human Factors in
  Computing Systems}, CHI '11, pages 2765--2768, New York, NY, USA, 2011. ACM.

\bibitem{tak2013satisficing}
Susanne Tak, Piet Westendorp, and Iris van Rooij.
\newblock Satisficing and the use of keyboard shortcuts: Being good enough is
  enough?
\newblock {\em Interacting with Computers}, 25(5):404--416, 2013.

\end{thebibliography}

\end{document}